\documentclass[final,twocolumn,prl,aps,floats,amsfonts,amssymb,stmaryrd,showpacs]{revtex4}

\usepackage{amsmath}
\usepackage{graphicx}
\usepackage{bm}

\def\EQ{\begin{equation}}
\def\EN{\end{equation}}
\def\EQA{\begin{eqnarray}}
\def\ENA{\end{eqnarray}}

\begin{document}

\title{Influence of turbulence on the dynamo threshold}
\author{J-P. Laval$^1$, P. Blaineau$^2$, N.
Leprovost$^2$, B. Dubrulle$^2$ and F. Daviaud$^2$}

\affiliation{$^1$ Laboratoire de M\'ecanique de Lille, CNRS, UMR 8107, Blv
Paul Langevin, F-59655 Villeneuve d'Ascq Cedex, France\\
$^2$ SPEC/DRECAM/DSM/CEA Saclay and CNRS,
URA2464, F-91190 Gif sur Yvette Cedex, France}

\begin{abstract}
We use direct and stochastic numerical simulations
of the magnetohydrodynamic equations to
explore the influence of turbulence on the dynamo
threshold. In the spirit of the
Kraichnan-Kazantsev model, we model the
turbulence by a noise, with given amplitude,
injection scale and correlation time.  The
addition of a stochastic noise to the mean
velocity significantly alters the dynamo
threshold. When the noise is at small (resp.
large) scale, the dynamo threshold is decreased
(resp. increased). For a large scale noise, a finite
noise correlation time reinforces this effect.
\end{abstract}


\maketitle

\vspace{0.1cm}

The process of magnetic field generation through
the movement of an electrically conducting medium
is called a dynamo. When this medium is a fluid,
the instability results from a competition
between magnetic field amplification via
stretching and folding, and damping through
magnetic diffusion. This is quantified by the
magnetic Reynolds number $Rm$, which must exceed
some critical value $Rm_c$ for the instability to
operate. Despite their obvious relevance in natural
objects, such as stars, planets or galaxies,
dynamos are not so easy to study or model.
Computer resources limit the numerical study of
dynamos to a range of either small Reynolds
numbers $Re$ (laminar dynamo), modest
$Rm$ and $Re$ \cite{Glatzmaier95} or small $Pm=Rm/Re$ using Large 
Eddy Simulation \cite{Ponty04}. These difficulties explain the recent 
development
of experiments involving liquid metals, as a way
to study the dynamo problem at large Reynolds
number. In this case, the flow has a non-zero
mean component and is fully turbulent. There is,
in general, no exact analytical or numerical
predictions regarding the dynamo threshold.
However, prediction for the mean flow action can
be obtained in the so-called "kinematic regime"
where the magnetic field back reaction onto the flow is neglected  (see
e.g. \cite{Dudley89}). This approximation is very
useful when conducting optimization of
experiments, so as to get the lowest threshold
for dynamo action based only on the mean flow
$Rm_{c}^{MF}$
\cite{Stefani99,Marie03,Forest02,Ravelet05}. It
led to very good estimate of the measured dynamo
threshold in the case of experiments in
constrained geometries
\cite{Gailitis00}, where the
instantaneous velocity field is very close to its
time-average.\

In contrast, unconstrained experiments
\cite{Peffley00,Forest02} are
characterized by large velocity fluctuations, allowing the exploration of
the influence of turbulence onto the mean-flow
dynamo threshold. Theoretical predictions regarding this influence 
are scarce. Small velocity fluctuations produce little impact on the 
dynamo threshold \cite{Petrelis02}. Predictions for arbitrary 
fluctuation amplitudes can be reached by considering the turbulent 
dynamo as
an instability (driven by the mean flow) in the
presence of a multiplicative noise (turbulent
fluctuations) \cite{LeprDubr05}. In this context, fluctuations favor 
or impede the magnetic
field growth depending on their
intensity  or correlation time. This observation
is confirmed by recent numerical simulations of
simple periodic flows with non-zero mean flow
\cite{Ponty05,Bayliss05} showing that
turbulence increases the dynamo threshold.\

In the sequel we use direct and stochastic
numerical simulation of the magnetohydrodynamic
(MHD) equations to explore a possible explanation,
linked with the existence of non-stationarity of
the largest scales. We found that the addition of
a stochastic noise to the mean velocity could
significantly alter the dynamo threshold. When
the noise is at small scale, the dynamo threshold
is decreased, while it is increased for a large
scale noise. In the latter case, the noise
correlation time plays a role, and reinforces
this effect, as soon as it is larger than the
mean eddy turnover time. When interpreted within
the Kraichnan-Kazantsev model of MHD flow, these
results predict that large scale (resp. small scale) turbulence
inhibits (resp. favors) dynamo action.

The MHD equations for incompressible fluids are :
\EQA
\partial_t {\bf u} +{\bf u}\cdot \nabla {\bf u}
&=& -\nabla P + \nu \nabla^2 {\bf u}+{\bf
j}\times {\bf B}+f(t) {\bf v}^{TG},\nonumber\\
\partial_t {\bf B}& =& \nabla \times\left({\bf
u}\times {\bf B}\right)+\eta \nabla^2 {\bf B}.
\label{MHD}
\ENA
Here, ${\bf u}$ is the velocity, ${\bf B}$ is the
Alfven velocity, $P$ the pressure, $\nu$ the
viscosity, $\eta$ the magnetic diffusivity,
${\bf j}=\nabla\times({\bf B})$, and ${\bf
v}^{TG}=(\sin x\cos y\cos z,-\cos x\sin y\cos z,0)$
is the Taylor-Green vortex and $f(t)$ is set by
the condition that the (1,1,1) Fourier components
of the velocity remains equal to ${\bf v}^{TG}$.
The equations are integrated on a triply periodic
cubic domain using a  pseudo-spectral method. The
aliasing is removed by setting the solution of
the 1/3 largest modes to zero. The time marching
is done using a second-order finite difference
scheme. An Adams-Bashforth scheme is used for the
nonlinear terms while the dissipative terms are
integrated exactly. The two control parameters
are the Reynolds number $Re=v_{rms}l_{int}/\nu$
and the magnetic Reynolds number
$Rm=v_{rms}l_{int}/\eta$, where
$v_{rms}=(1/3)\sqrt{2 E}=(1/3)\sqrt{<u^2>}$ is
the (spatial) r.m.s. velocity
based on the total kinetic energy $E=\int E(k)
dk$ and $l_{int}=(3\pi/4)E/\int k E(k) dk$ is the
integral scale of the turbulent flow. Both
$v_{rms}$ and $l_{int}$ fluctuate with time.
Thus, viscosity and diffusivity are dynamically
monitored so as to keep $Re$ and $Rm$ constant.
We have checked that $Re$ is a simple linear
function of a non-dynamical Reynolds number
$Re_{\rm exp}= v_{\rm max} \pi/\nu$ (usually used in
experiments) based on maximum velocity and half
the simulation box: $Re=7.41 Re_{\rm exp} $. In the
sequel $<X>$ (resp. $\overline{X}$) refers
to spatial (resp. time) average of $X$.\

We ran typically four types of simulations : i)
DNS-MHD, where the full set of equation
(\ref{MHD}) is integrated at $5\le Re\le 100$ and
$5\le Rm\le 50$ using resolutions up to $256^3$ ;
ii) LES-MHD, where the Lesieur-Chollet model of
turbulence is used for the velocity equation
(\ref{MHD}-a), allowing to explore
a case out-of-reach of DNS \cite{Ponty05}, namely
   $Re=500$, $5\le Rm\le 100$; iii)
kinematic simulations ; iv) kinematic-stochastic (KS)
simulations. In these last two cases, only the
induction equation (\ref{MHD}-b) is integrated
with ${\bf u}$ set to a given velocity field. In
the kinematic case, it corresponds to
the stationary velocity
field $\overline{\bf u}(Re)$ obtained through time-average of a stable
solution of the Navier-Stokes equations with
Taylor-Green forcing, at fixed Reynolds number.
This procedure is complicated by the presence of
hydrodynamic instabilities at low Reynolds number
\cite{LDDD05}, which impose very long simulation
time (typically over {1000 s}, i.e. 400 eddy turn
over times at $Re=46$) to ensure convergence towards an
asymptotically stable solution. The average is
then performed over several (typically 200)
eddy-turnover times. In the KS
case, the velocity field ${\bf u}=\overline{\bf
u}(Re)+{\bf v'}(k_I,\tau_c)$ is the sum of a time
averaged velocity field at a given $Re$ and of an
external Markovian Gaussian noise, with fixed
amplitude $v'$, correlation time $\tau_c$ and
typical scale $k_I$. In both kinematic
simulations, the magnetic Reynolds number $Rm$ is
computed by using the rms velocity and integral
scale of ${\bf u}$. In the deterministic case,
this amounts to use
$V_{rms}=(1/3)\sqrt{<\overline{u}^2>}$
and $L_{int}$ the (spatial) rms
velocity and integral scale of the time-averaged
velocity field, therefore respecting the
experimental procedure followed in optimization
of dynamo experiments \cite{Stefani99,Marie03,Ravelet05,Forest02}.\

For each type of simulation, we fix $Re$ (
$v'$, $\tau_c$ and $k_I$, if needed), vary $Rm$
and monitor the time behavior of the magnetic
energy $<B^2>$ and the finite-time Lyapunov
exponent $\Lambda=0.5\ \partial_t <\ln(B^2)>$,
where the average is taken over the spatial
domain. Three types of behaviors are typically
observed \cite{LDDD05}: i) no dynamo : the
magnetic energy decays, the Lyapunov converges
towards a finite negative value; ii) undecided
state : the magnetic energy remains at a low
level, with intermittent bursts of magnetic
energy \cite{Ott01}  and oscillation of the Lyapunov, so that
no fit of the Lyapunov exponent can be obtained;
iii) turbulent dynamo : the magnetic energy grows
with positive Lyapunov, and, in the DNS-MHD or
LES-MHD, reaches a nonlinear saturated regime.\

 From the values of the Lyapunov in the turbulent
dynamo and no dynamo regime, one may derive the
critical magnetic Reynolds number $ Rm_{c}(Re)$,
solution of  $\Lambda(Re, Rm_c)=0$, through a
standard interpolation procedure.
\begin{figure}[h]
\begin{center}
\includegraphics[scale=0.45,clip]{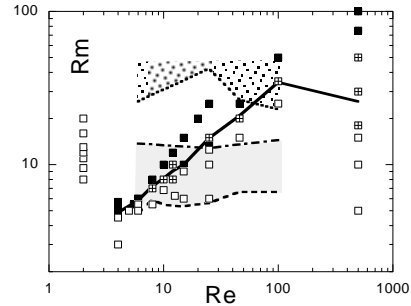}
\caption{Simulation parameter space. Square refer
to DNS-MHD and LES-MHD simulations, and shaded
areas to windows of dynamo action for kinematic
simulations with mean flow. $\square$ : no-dynamo
case ; $\boxplus$ : undecided state;
$\blacksquare$ : dynamo case ;  $-$ $Rm_c^{turb}$ ; $---$ $Rm_c^{MF}$ 
; $-\cdot-\cdot$ end of the first kinematic dynamo window ; 
$\cdot\cdot\cdot$ beginning of the second kinematic dynamo window. 
Shaded areas indicate explored windows of dynamo action for the mean 
flow. }
\label{figure-recap}
\end{center}
\end{figure}

A summary of our exploration of the parameter
space is provided in Fig. 1, for the
non-stochastic simulations, where the only
control parameters are $Rm$ and $Re$. We did not detect any dynamo at 
$Re=2$. Between $Re=4$ and $Re=6$, we observed heterocline dynamos, 
oscillating between a non-dynamo and a dynamo state. The window 
$2<Re<4$ has been studied in \cite{Ponty05}, where decreasing 
critical magnetic Reynolds number has been found.  For $4<Re<100$, we 
found that the critical magnetic Reynolds number for
dynamo action in a turbulent flow $Rm_c^{turb}$
increases with the Reynolds number, in
quantitative agreement with the result
obtained in the same geometry, but with a
different forcing (at constant force instead of constant velocity) 
\cite{Ponty05}. Our LES-MHD simulation confirms the saturation of the 
dynamo threshold at large Reynolds number already observed in 
constant force simulations \cite{Ponty05}. For the mean
flow, we have actually detected at least two
windows of dynamo actions : one, independent of
$Re$, starting above $Rm_c^{MF}\approx 6$ and
centered around $Rm=10$, with real Lyapunov
(stationary dynamo); a second, occurring at larger
$Rm$, varying with $Re$, with complex Lyapunov
(oscillatory dynamo). One sees that $Rm_c^{turb}$
varies across these two windows and always
exceeds $Rm_c^{MF}$. In the sequel, we show that
the increase and saturation of $Rm_c^{turb}$ is not due to a crossing 
between the two dynamo modes, but to the influence of non-stationary 
large scales over $Rm_c^{MF}$.\

\begin{figure}[h]
\begin{center}
\begin{minipage}{\columnwidth}
\includegraphics[width=0.49\columnwidth]{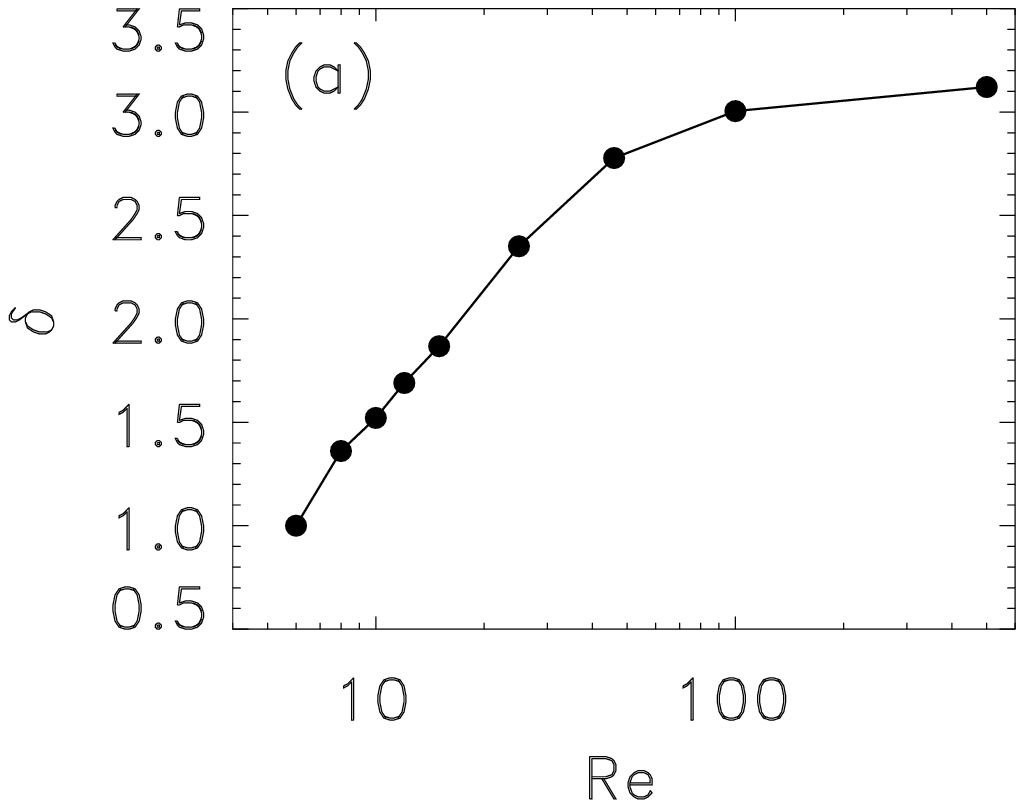}
\includegraphics[width=0.49\columnwidth]{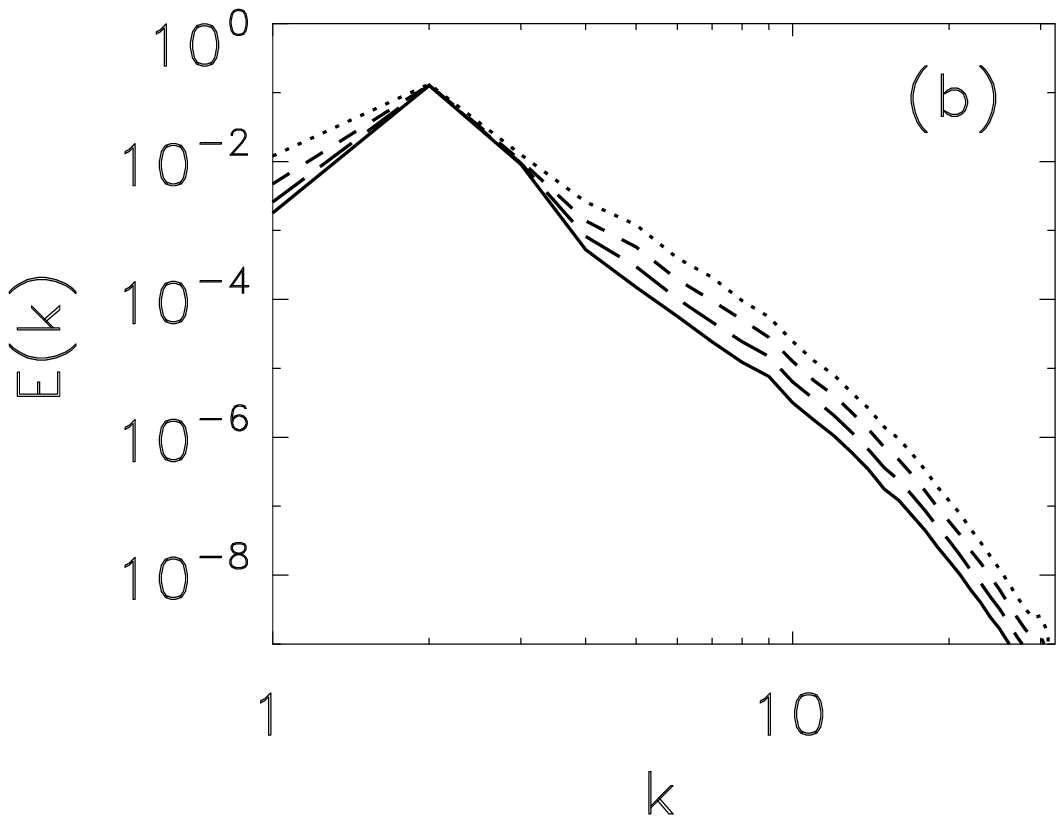}
\end{minipage}
\caption{a) Noise intensity
$\delta  = \overline{<u^2>}/<\overline{u}^2>$, as a function of
the Reynolds number, measured in our DNS
simulations under the dynamo threshold. b) Energy
spectrum of the velocity field in the DNS, at
$Re=46$, for different average period $T$: dotted
line : $T=0$ ; short dashed line : $T=75$ s ;
long-dashed line : $T=150$ s ; continuous line :
$T=300$ s. }
\label{noise}
\end{center}
\end{figure}

To make an easier connection between DNS and KS
  simulations, we introduce a parameter that quantifies
the noise intensity, $\delta  = \overline{<u^2>}/<\overline{u}^2>$.
This parameter depends on the noise amplitude, as
well as its correlation time and characteristic
scale, and need to be computed for each
stochastic simulation. It can also be computed in
the direct simulations, and is found to depend
onto the Reynolds number, increasing from a value
of $1$ at low Reynolds number, until about $3$ at
the largest available Reynolds number (Figure
\ref{noise}-a). Note that $\delta-1$ is just the
ratio of the kinetic energy of fluctuations onto
the kinetic energy of the mean flow. In the
sequel, the comparison between the
KS and DNS-MHD simulations
will therefore be made using $\delta$ as the
control parameter. Another interesting
information can be obtained from the energy
spectrum of the velocity field, as one averages
over longer and longer time scales (Figure
\ref{noise}-b). One sees that during the first
period of average (typically, a few eddy
turn-over time, i.e. about $5$ to $10$ s), one
mainly removes the fluctuations at largest
scales, while the remaining average mostly
removes small scales (over time scales of the
order of $50$ to $100$ eddy-turnover times, i.e.
$300$ s).\

\begin{figure}[h]
\begin{center}
\begin{minipage}{0.49\columnwidth}
\includegraphics[width=\columnwidth]{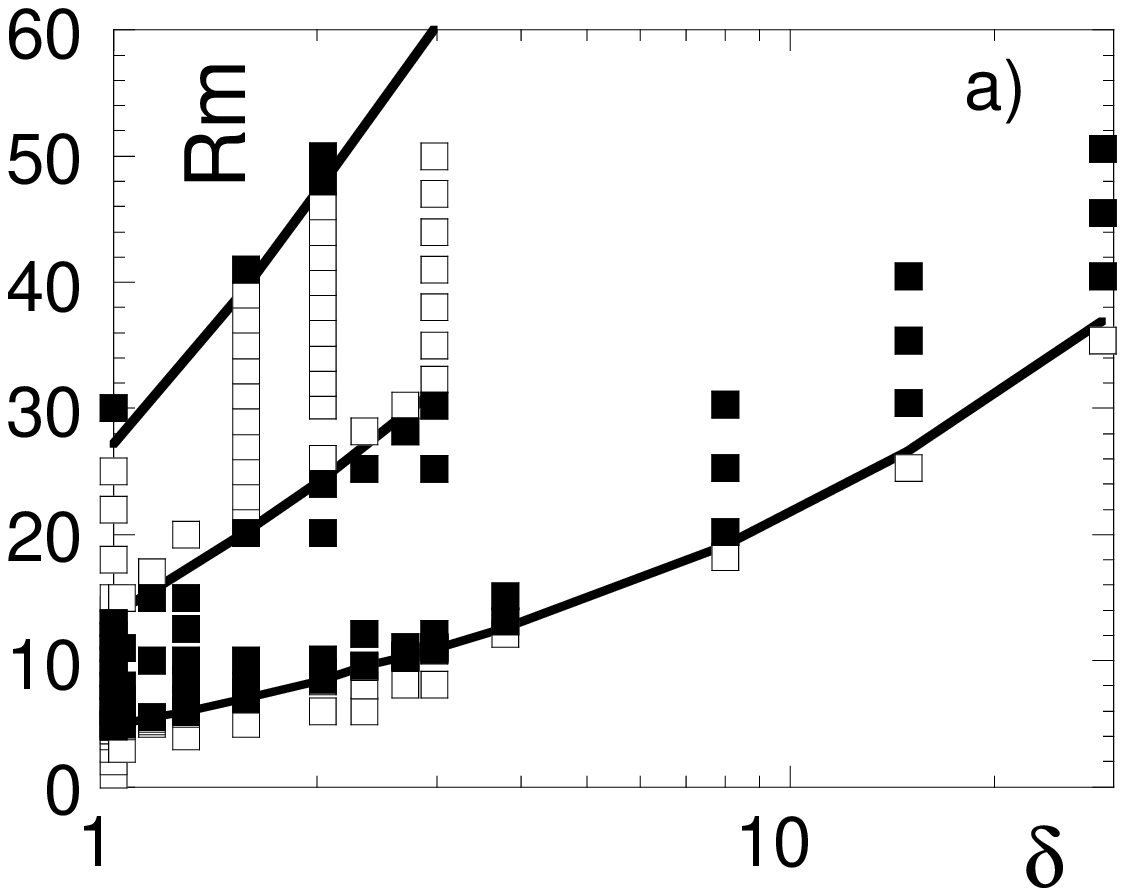}
\includegraphics[width=\columnwidth]{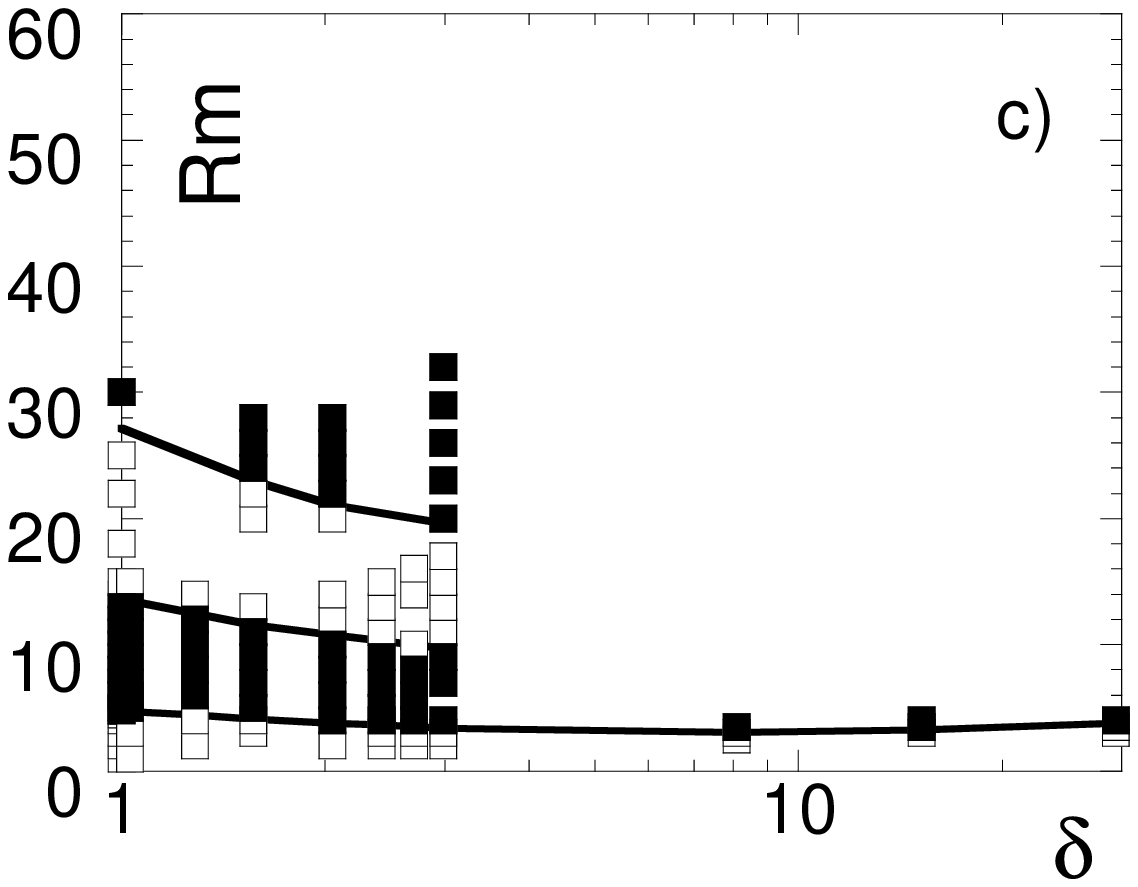}
\end{minipage}
\hfill
\begin{minipage}{0.49\columnwidth}
\includegraphics[width=\columnwidth]{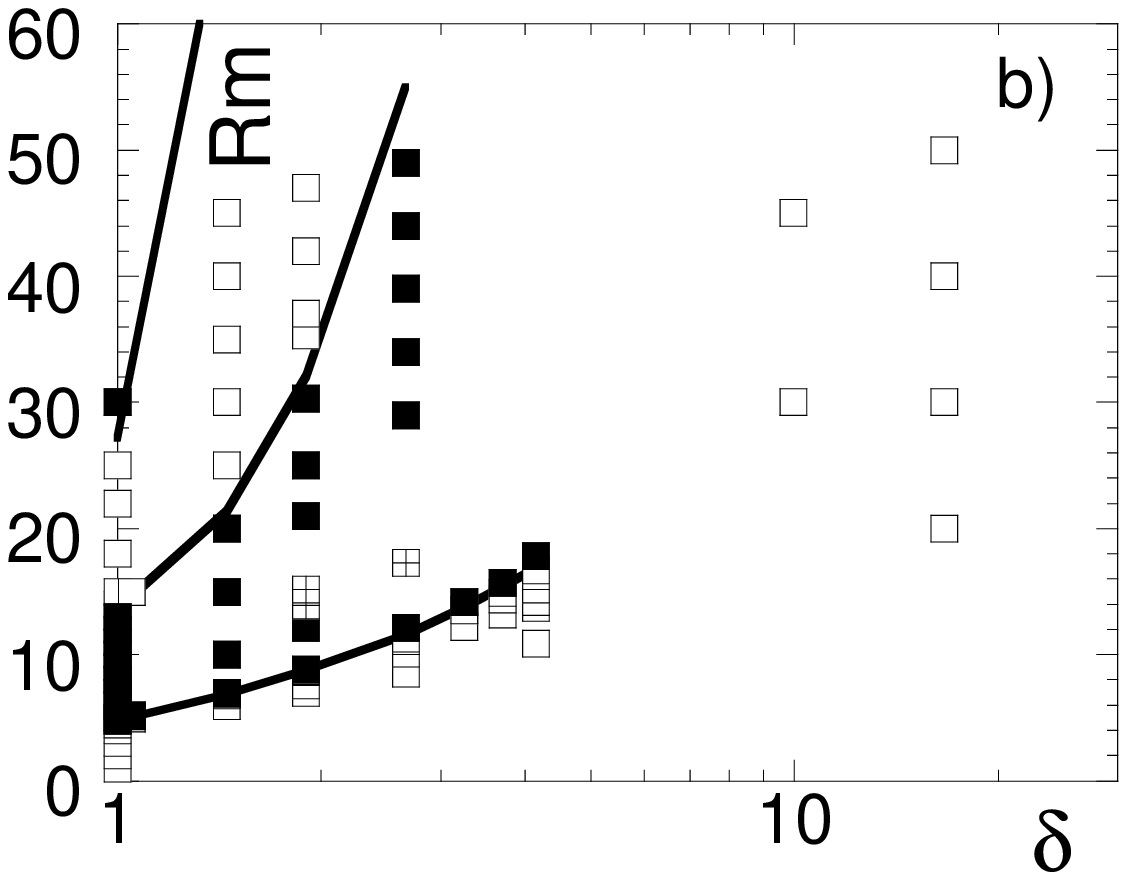}
\includegraphics[width=\columnwidth]{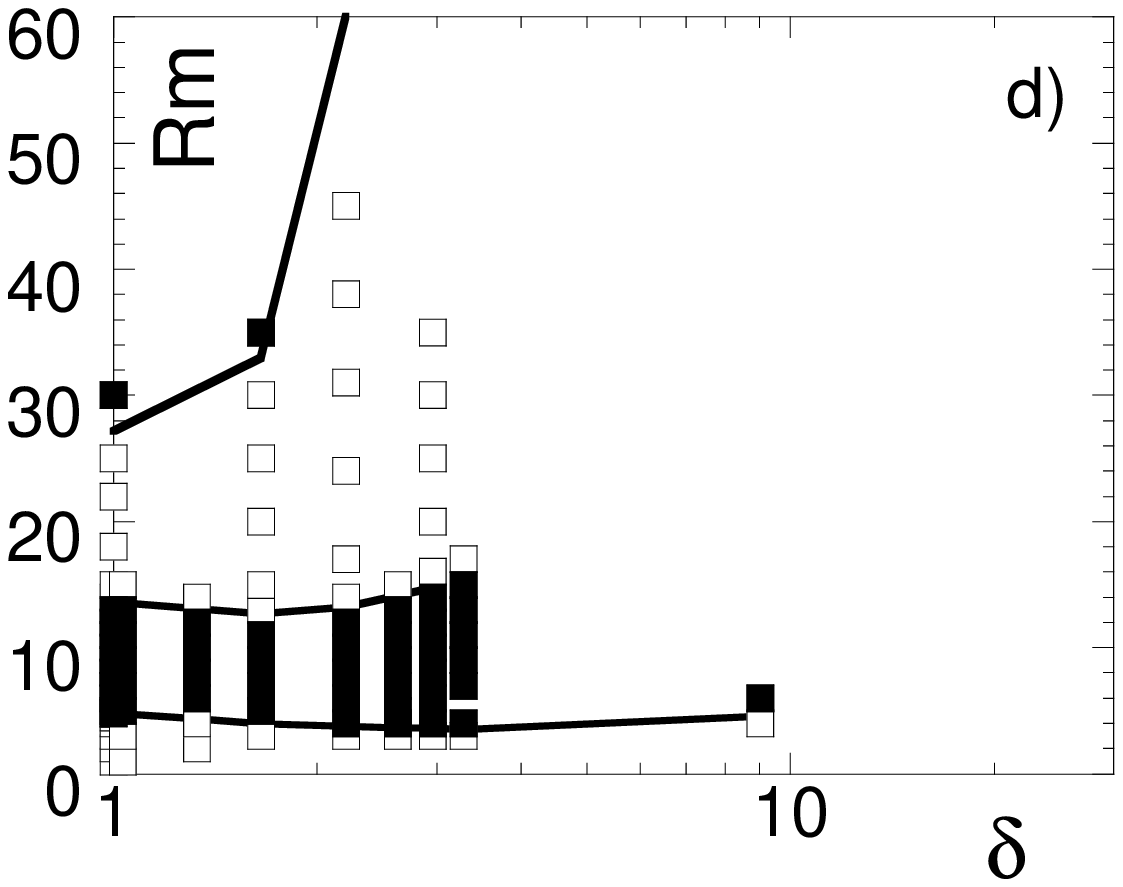}
\end{minipage}
\caption{Parameter space for noise at $Re=6$ for
different noise parameters : a) $\tau_c=0$,
$k_I=1$; b) $\tau_c=0.1$ sec, $k_I=1$; c) $\tau_c=0$,
$k_I=16$; d) $\tau_c=0.1$ sec, $k_I=16$. $\square$ :
no-dynamo case; $\boxplus$ : undecided state;
$\blacksquare$ : dynamo case. The full lines are
zero-Lyapunov lines.}
\label{Fig-noise6}
\end{center}
\end{figure}

\begin{figure}[h]
\begin{center}
\begin{minipage}{0.49\columnwidth}
\includegraphics[width=\columnwidth]{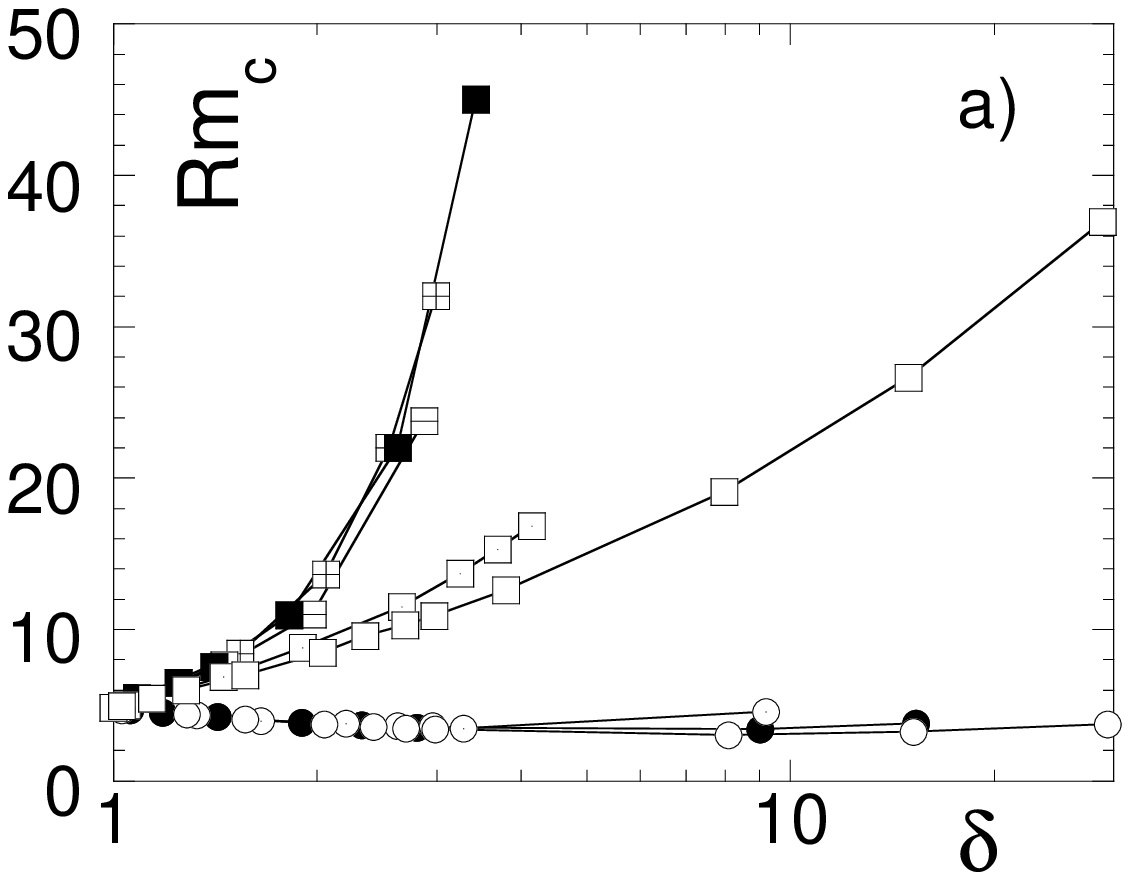}
\end{minipage}
\begin{minipage}{0.49\columnwidth}
\includegraphics[width=\columnwidth]{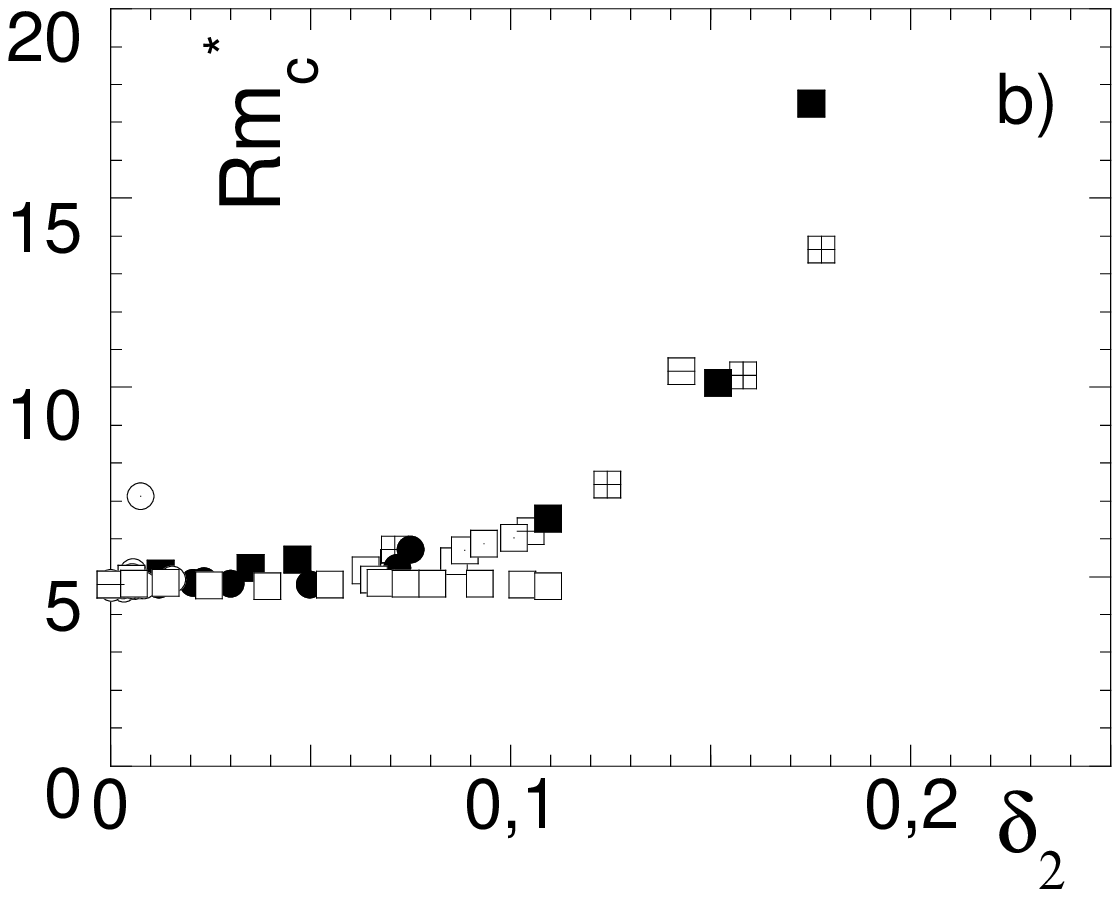}
\end{minipage}
\caption{Evolution of the dynamo threshold for
KS simulations with
$\overline{u}(Re=6)$. a) $Rm_c$ as a
function of $\delta$ and  b) $Rm_c^*$ as a function
of $\delta_2$ for different noise parameters :
$k=1$ : $\square$
$\tau_c=0$ ; $\boxdot$ $\tau_c=0.1$ sec; $\boxminus$
$\tau_c=1$ sec; $\boxplus$ $\tau_c=8$ sec ;
$\blacksquare$ $\tau_c=50$ sec ; $k=16$ : $\circ$
$\tau_c=0$ ; $\odot$
$\tau_c=0.1$ sec; $\bullet$ $\tau_c=50$ sec.}
\label{Fig-nois-sum}
\end{center}
\end{figure}

In the sequel, we
explore the influence of both type of
fluctuations through the
KS simulations, by considering noise at large
($k_I=1$) and small scale ($k_I=16$), with
correlation time ranging from $0$ to $50$ s.
Since the
kinematic dynamo threshold is essentially
constant for all values of the Reynolds number we
explored, we first focus on the study of the case
where the time-averaged field is fixed as
${\overline u}(Re=6)$ and vary the noise
amplitude, characteristic scale or correlation
time, to explore their influence on the dynamo
threshold. An example of our exploration of the parameter
space is provided in Figure
\ref{Fig-noise6}, for different kinds of noise
and $\overline{u}(Re=6)$.
Note that by using our external noise, we are
able to produce noise intensities of the order of
noise intensities measured in experiments
($\delta\sim 10$ at $Re\sim 10^6$ for the von Karman flow), and that are out of
reach of DNS. For low correlation time or
injection scale, we are actually able to follow
the deformation of the two windows of dynamo
action. One sees that a noise does not destroy
them, but rather distorts them. In the case where
the noise is at small scale ($k_I=16$), the two
windows are slightly tilted downwards, while they
are lifted upwards in the case of large scale
noise ($k_I=1$). The influence of the noise onto
the first dynamo bifurcation (the dynamo
threshold) can be summarized by plotting the
critical magnetic Reynolds numbers as a function
of the noise intensity (Fig.
\ref{Fig-nois-sum}-a). Large scale (resp. small-scale)
noise tends to increase (resp. decrease) the dynamo
threshold. Furthermore, one
sees that for small scale noise, the decrease in
the dynamo threshold is almost independent of the
noise correlation time $\tau_c$, while for the large scale
noise, the increase is proportional to $\tau_c$ at small $\tau_c$.
At $\tau_c\agt 1$ sec-one third of
the mean eddy-turnover time-, all curves $Rm_c(\delta)$ collapse onto 
the same curve.  We have further
investigated  this behavior to understand
its origin. Increasing $\delta$ first increases of the flow « turbulent
viscosity » $\overline{v_{rms}}
\overline{l_{int}}$ with respect to its mean flow value 
$V_{rms}L_{int}$. This effect can be corrected by considering 
$Rm_c^*=Rm_c V_{rms}
L_{int}/ \overline{v_{rms}}\overline{ l_{int}}$.
Second, an increase of $\delta$ produces an increase of the fluctuations of
kinetic energy, quantified by 
$\delta_2=\sqrt{\overline{<u^2>^2}-\overline{<u^2>}^2}/\overline{<u^2>}$. 
This last effect is more pronounced at
$k_I=1$ than at $k_I=16$. It is amplified through
increasing noise correlation time. We thus
re-analyzed our data by plotting $Rm_c^\ast$
as a
function of $\delta_2$
  (Fig. \ref{Fig-nois-sum}-b). All results tend to collapse onto a single
curve, independently of the noise injection scale
and correlation time. This curve tends to a
constant equal to $Rm_c^{MF}$ at low $\delta_2$.
This means that the magnetic diffusivity needed to
achieved dynamo action in the mean flow is not
affected by spatial velocity fluctuations. This is achieved for small scale
noise, or large scale noise with small
correlation time scale. In contrast, the curve
diverges for $\delta_2$ of the order of $0.2$,
meaning that time-fluctuations of the kinetic energy superseding $20$ 
percent of the total
energy annihilate the dynamo.

\begin{figure}[h]
\begin{center}
\includegraphics[scale=0.45,clip]{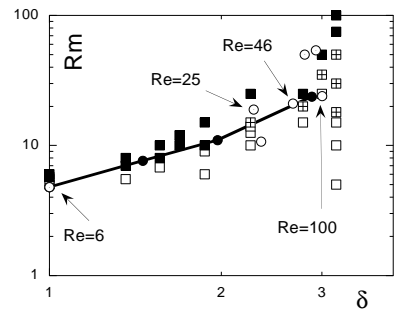}
\caption{Evolution of the dynamo threshold as a
function of $\delta$ for the DNS (squares, same
meaning as in Fig. 1 and 3) and
KS simulations with $k_I=1$ using $\overline{u}(Re=6)$, $\tau_c=1$ sec
($\bullet$ connected with line), and
$\overline{u}$ at the Reynolds number
corresponding to the DNS with equivalent $\delta$
($\circ$ for $\tau_c=1$ sec and $\odot$ for $\tau_c=8$ sec).}
\label{Fig-nois-DNS}
\end{center}
\end{figure}

We now turn to detailed comparison of dynamo thresholds
obtained in KS simulation
with DNS-MHD case. In Fig. \ref{Fig-nois-DNS}, we show that the 
dynamo threshold
obtained at $k_I=1$, for $\tau_c\agt 1$ as a
function of the noise intensity corresponds to
the DNS-MHD dynamo threshold. Note that the noise
  intensity $\delta$ saturates past
a Reynolds number of about $100$ (Fig. \ref{noise}-a), thereby
inducing the saturation of the
critical magnetic Reynolds number at large Reynolds number. To check that
our results are not affected by the choice of
$\overline{u}$, we ran additional KS
simulations with $\overline{u}$ computed at
$Re=25,46$ and $100$. Since the computational
cost in these cases is much larger than in the
case $Re=6$, we focused on the case where the
noise has a correlation time $\tau_c=1$ or $8$ sec. and
injection scale $k_I=1$ and only computed the
critical magnetic Reynolds number for the level
of noise reached by the DNS at that Reynolds
number. Figure
\ref{Fig-nois-DNS} shows that the dynamo
threshold coincides with the dynamo thresholds both  of KS 
simulations at $Re=6$
and of the DNS, indicating that a large scale
noise is probably responsible from the increase
of $Rm_c^{turb}$ with Reynolds number. A physical
identification of the dynamics of the velocity
fluctuations playing the role of this noise can be
performed by visual inspection of the turbulent
velocity field. One observes that the large scale
vortices generated by the Taylor-Green forcing
are not exactly stationary, but wander slightly
with time. A similar large-scale non-stationarity has been observed 
in the shear layer of Von Karman flows\cite{Ravelet05b,Volk05}. This 
process is approximately reproduced by a large
scale noise with sufficiently long correlation
time, and may therefore been held responsible for
the significant increase of the dynamo threshold.

Our work suggests that it might not be so easy to
achieve turbulent dynamos in unconstrained
geometries, with large scale non-stationarity. In
the experiments, a necessary ingredient for dynamo action could 
therefore be a monitoring
of the large scale, so as to keep them as
stationary as possible. In geo- and astrophysical
flows, this role could be played by the Coriolis
force. Our work
also indicates that a well chosen noise can be
used in place of the actual turbulent velocity
fluctuations to compute the dynamo threshold, at
a much lower computational cost. In some sense, a
kinematic-stochastic simulation can therefore be
seen as a turbulent model and might be useful in
the astro- or geophysical context.

{\bf Acknowledgments}\
We thank the  GDR Turbulence and GDR Dynamo for
support, J-F. Pinton, Y. Ponty, A. Chiffaudel and F. Plunian for
discussions, and E. Gouillard for logistical help
and CPU. Numerical simulations were performed at
IDRIS.

\end{document}